\documentclass[aps,pre,groupedaddress,showpacs,a4paper]{revtex4}

\input{epsf}
\usepackage{feynmf}

\usepackage{amssymb,amsfonts,amsmath,color}
\usepackage{graphicx}
\usepackage{pdfpages}

\begin{document}



\title{ Glassy Critical Points and the Random Field Ising Model }




\author{Silvio Franz(1), Giorgio Parisi(2) and Federico Ricci-Tersenghi (2)
}

\affiliation{
{\small (1) Laboratoire de Physique Th\'eorique et Mod\`eles  Statistiques,}\\
{\small CNRS et Universit\'e Paris-Sud 11, UMR8626, B\^at. 100, 91405 Orsay  Cedex, France}\\
{\small (2) Dipartimento di Fisica, INFN -- Sezione di Roma I, IPCF-CNR -- UOS Roma, Sapienza Universit\`a di Roma, P.le Aldo Moro 2, 00185 Roma, Italy}
}

\date{\today}

\begin{abstract}
  We consider the critical properties of points of continuous glass
  transition as one can find in liquids in presence of constraints or
  in liquids in porous media. Through a one loop analysis we show that
  the critical Replica Field Theory describing these points can be
  mapped in the $\phi^4$-Random Field Ising Model. We confirm our
  analysis studying the finite size scaling of the $p$-spin model defined
  on sparse random graph, where a fraction of variables is frozen such
  that the phase transition is of a continuous kind.
\end{abstract}

\maketitle

\section{Introduction}

The last years of research have emphasized the importance of
fluctuations in understanding glassy phenomena. The present
comprehension of long lived dynamical heterogeneities in supercooled
liquids compares the growth of their typical size to the appearance of
long range correlations at second order phase transition points
\cite{het}. Unfortunately, in supercooled liquids, the theoretical
study of these correlations beyond the Mean Field is just at an
embryonic level. It has been recently proposed that the putative
discontinuous dynamical transition of Mode Coupling Theory, which is
present when all activated processes are neglected, belongs to the universality
class of the unstable $\phi^3$ theory in a random field ($\phi^3$-RFIM
in the following) \cite{gangof4}. However in real
systems the activated processes cannot be neglected, the only remnant
 of the transition is a dynamical crossover and it is not 
clear if there is a range where the prediction of the theory can be tested.

In usual phase transition often we have a line of first order
transitions that ends at a second order terminal critical point. The
most popular case are ferromagnets: at low temperatures there is a
first order transition when the magnetic field crosses zero (the
magnetization has a discontinuity) and this transition lines end at
the usual critical point. The same phenomenon happen for the gas
liquid transition: it is a first order transition at low temperatures
that ends in a second order transition at the critical point.

A similar situation can occur for liquids undergoing a glass
transition, where lines of discontinuous glass transitions can
terminate in critical points.  In this note we focus our attention to
these terminal points, where the glass transition becomes continuous
and activation does not play a major role in establishing
equilibrium. This transition has both a dynamic and a thermodynamic
character, and it is not necessarily wiped out in finite dimension.
Glassy critical points have been theoretically studied in detail both
at the dynamic and at the thermodynamic level. In dynamical Mode
Coupling Theory (MCT)\cite{goetze} these are known as $A_3$
singularities, and have been recently observed in simulations of
kinetically constrained models on Bethe lattice \cite{sellitto}. At
the thermodynamic level they are known from mean field Spin Glass
models \cite{pspincampo} and Integral Equations approximations of
liquid theory \cite{boh}.  At these points the discontinuity in the
Edwards-Anderson non-ergodicity parameter vanishes and
correspondingly, the separation of dynamics in alpha and beta regime
is blurred.  On approaching the critical points from the discontinuous
transition side the MCT exponents characterizing the beta relaxation
go to zero and the alpha relaxation follows a universal scaling
function \cite{a3}.
In general bulk liquids display a discontinuous transition pattern.
However, the transition can become continuous for a particular 
choice of the parameters, e.g. in presence of constraints or 
of quenched disorder.
It has been argued that within MCT the glass transition
can become continuous for liquids are confined in porous media 
\cite{krackoviac}. If
one studies the transition as a function of the spatial 
density of the
confining matrix $\rho_M$, one finds lines of discontinuous dynamic 
and thermodynamic transition that get displaced 
at lower and lower liquid densities, until they merge at a common 
critical point where the transition becomes continuous. 

From the theoretical side a suitable way of constraining a glassy
system consists in introducing a ``pinning field'' term in the
Hamiltonian pushing the system in the direction of a randomly chosen
reference equilibrium configuration. In \cite{FPprl} it was proposed a
phase diagram in the plane of temperature and pinning field, showing
lines of first order dynamical and thermodynamical transition that
merge and terminate in a common critical point as reproduced in figure
1.  This view, based on simple spin glass models, was confirmed for
liquids in the replica hypernetted chain approximation in \cite{hnc}
and supported by numerical simulation of realistic model liquids in
\cite{FPphysicaA,hnc}.

\begin{figure}
\includegraphics{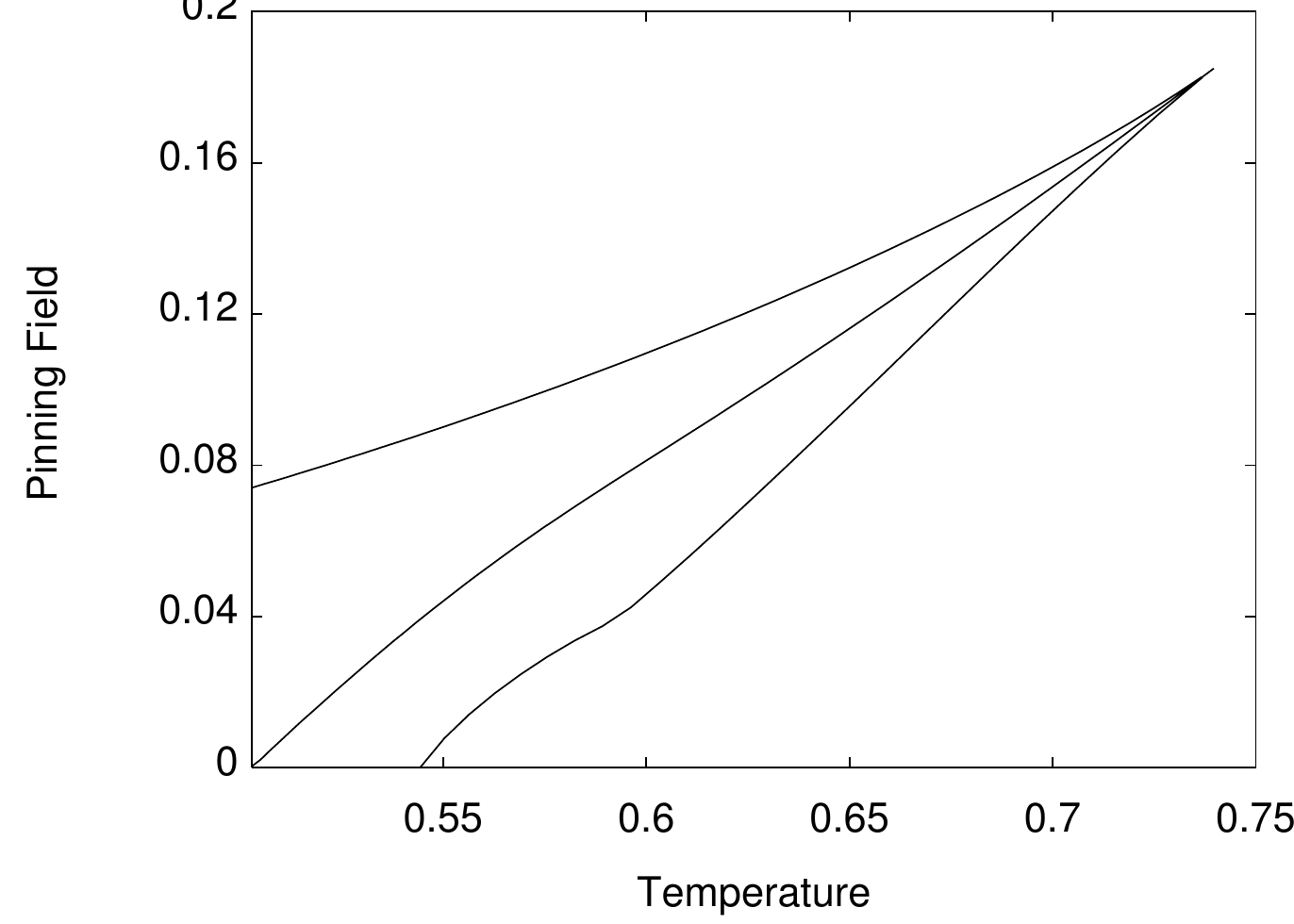}
\caption[0]{\protect\label{F_E1} Phase diagram in the plane of Temperature and 
Pinning Field in a schematic mean field system. 
At small temperatures for large values of the pinning field the system  at equilibrium stays near the reference configuration (high overlap phase) while at small values of the pinning field
the system   stays far from the reference configuration (low overlap phase).
The upper curve is line of dynamical transition below which 
a dynamically stable high overlap phase exist.
The middle curve is the thermodynamic first order
transition line between the low overlap and the high overlap phases. 
The lower curve is the spinodal of the high overlap phase, i.e. the point where it becomes unstable. 
The three curves terminate in the critical terminal point. (Figure from: S. Franz and G.Parisi, PRL {\bf 79} 2486 (1997)). }
\end{figure}  

More recently, the interest for the phase diagram of constrained
systems has been renewed by liquid simulations where a finite fraction
of the particles are frozen to the position they take in a selected
equilibrium configuration\cite{BK}. The effect of the frozen particles
is similar to the one of an adsorbing matrix in a porous medium or of
a pinning field, with the important addition that in this case the
unfrozen particles remain in the original equilibrium state. 

The detailed phase diagram for spin models on random graphs was
computed in \cite{MRS,RTS}. A theoretical discussion of the physical
relevance of this situation for liquids and glasses and an exact
computation for the mean field $p$-spin models were presented in
\cite{BC}. Differently from the case of the pinning field
where the field transforms the glass transition into 
first order and spinodal transitions, in the case of frozen particles 
the dynamic and thermodynamic transition lines keep
with their glassy random first order character that one finds at zero pinning.  

In all these cases, the existence of a critical terminal  point is interesting
because while the dynamical critical line has to disappear in finite
dimension thanks to dynamical activation, the critical terminal  point, which is
also the terminating point of the thermodynamic transition line could
survive in finite dimensions and can be studied in numerical simulations and
experiments.

As it is usual for lines of phase transitions terminating in a
critical point, the critical terminal point lies in a different
universality class of the line. The critical properties of
discontinuous dynamical transitions has been recently analyzed in
\cite{gangof4}. It has been proposed that the time independent part of
the fluctuations in the $\beta$ and early $\alpha$ dynamical regimes
admit a description in terms of a cubic replica field theory. The
leading singularities of this theory in perturbation theory happen to
coincide with the ones of a $\phi^3$ field theory in a random magnetic
field. At the critical point the coefficient of the $\phi^3$ term
vanishes and it is natural to make the hypothesis that the next relevant
term is a $\phi^4$ term so that the resulting theory is the standard
$\phi^4$-RFIM \cite{RFIM}.

Arguments in this direction have been put forward in \cite{BC} using a
RG procedure. Unfortunately, the arguments in \cite{BC}, though
suggestive, are not fully convincing. They are based on a
Migdal-Kadanoff renormalization scheme, which uses a hybrid formalism
where replicas are used to average out the randomness in couplings,
but additional quenched disorder introduced to mimics the effect of
the frozen particle is kept unaveraged.

In this paper we use the tools of replica field theory to support the
hypothesis that glassy critical points are in the universality class
of the $\phi^4$-RFIM. We analyze in detail the case of a dynamical
transition line terminating in a critical point. Such a scenario
applies exactly in the case in which a finite fraction of particles
are pinned in an equilibrium condition. The case of a pinning field or
of an adsorbing matrix presents additional complications that will be
left to future work.  Replica field theory can be used to find out the
nature of this transition. We have to consider a system with $n$ clone
in presence of some constraint: when the number $n$ of components goes
to 1, one of the clones is at equilibrium and the other clones feel
the effect of a quenched field.  Since one does not specify which of
the clones is privileged the final theory is replica symmetric. From
a field theoretical perspective, at first site the RFIM hypothesis is
self-evident: indeed if we take care of the leading terms, the replica
theory corresponds to theory with a random temperature, that maps on a
theory with a potential $V(\phi)$ with a random magnetic field and the
critical terminal point is described by a $\phi^4$ interaction.
However this argument holds only for the leading terms and neglects
sub-leading orders that may play a crucial role if the leading terms
cancels.

More precisely the mapping of the dynamical
transition to the $\phi^3$-RFIM comes from the extraction of the most
singular contribution of a cubic replica field theory, with multiple
fields of different scaling dimension. The neglected sub-leading terms turn out
to become dimensionally relevant below dimension 6, the same dimension
as the quartic terms of the $\phi^4$-RFIM. The contribution of cubic
vertexes cannot therefore be directly dismissed on the basis of
dimensional analysis. To understand the nature of the terminal critical point
and the possibility of cancellations that restore the RFIM mapping, a
careful analysis of the perturbative series is needed. The scope of
this note is to investigate this problem at the level of Ginzburg
criterion, computing the one loop corrections to the propagators due
to the residual cubic vertexes and comparing them to the ones coming
from the quartic terms. We find that a-priori unexpected
cancellations are present so that the cubic vertexes contributions appear
to be irrelevant as compared to the quartic ones.

In order to test our theoretical results, we consider a diluted $p$-spin
model (random XOR-SAT problem) on a random graph in presence of frozen
spins and perform extensive numerical simulations to study finite size
scaling close to the critical point.  The simulations on large systems
fully confirm our analysis, making us confident that the one-loop
result indeed holds to all orders.

In the next section we present the theoretical analysis. In the
subsequent one the numerical simulation. We then conclude the
paper. An appendix is devoted to the technical details of the
theoretical computations.

\section{Analysis} 
Our starting point is the replica field theory \cite{DKT,CPR}
\begin{widetext}
\begin{equation}
{\mathcal L}={1 \over 2}\int dx \left( \sum_{ab} (\nabla
\phi_{ab})^2+m_1 \sum_{ab}\phi_{ab}^2+m_2\sum_{abc}
\phi_{ab}\phi_{ac}+m_3\sum_{abcd}\phi_{ab}\phi_{cd} \right) -{1 \over
  6}\omega_1 {\rm Tr\;} \phi^3-{1 \over 6}\omega_2
\sum_{ab}\phi_{ab}^3\;,
\label{rft}
\end{equation}
\end{widetext}
where the field $\phi_{ab}(x)$ is an $n\times n$ space dependent
matrix order parameter with vanishing diagonal terms, describing the
fluctuations of the correlation function around its plateau value. Among
all possible cubic replica invariant we have retained only the ones
giving rise to the leading and possible next to leading singular
behavior. It has been show in \cite{gangof4} that in the limit $n\to
1$ the theory describe fluctuations close to the dynamical transition,
that occurs for $m_1\to 0$ (while $m_2$ and $m_3$ remain finite).

According to the discussion in \cite{gangof4} the
components of the matrix $\phi_{ab}$ do not share a unique scaling
dimension. Similarly to what found in the RFIM problem  \cite{Cardy}, 
in order to use dimensional analysis one needs to change 
basis and define linear combinations $\phi$, $\omega$
and $\chi_{ab}$ of the $\phi_{ab}$ that exhibit good scaling
properties. The actual linear transformation reeds
\begin{eqnarray}
\phi_{ab}=(\phi-\frac 1 2 \omega)(1-\delta_{ab}) + U_{ab} \omega +\chi_{ab}
\label{trasf}
\end{eqnarray}
where $U_{ab}= \delta_{a,a-(-1)^a}$ is a size two band matrix and
  $\chi_{ab}$ is a symmetric matrix null on the diagonal and such that
\begin{eqnarray}
\sum_{b}\chi_{ab}=0 \;\;\; {\rm for} \;\;{\rm all} \;\;\; a
\nonumber
\\
\sum_{ab}U_{ab}\chi_{ab}=0.
\label{constraints}
\end{eqnarray}
The analysis of the resulting quadratic form (see below) 
shows that 
$\phi$, $\omega$
and $\chi_{ab}$ have well defined scaling dimension $D_\phi$,
$D_\omega$ and $D_\chi$ that, 
fixing the dimension $D_{m_1}=1$, verify $D_\omega=D_\phi+1$ and
$D_\chi=D_\phi+ \frac 1 2$. Fixing the dimension of $\phi$ by the
condition that the action is adimensional one finds $D_\phi=\frac D
4-1$. Notice that if the normalization of the dimension would be fixed
by the condition that the dimension of the momentum is one, we
should multiply all dimensions by a factor two.

  In terms of the new fields, keeping only the terms giving rise to
  the leading singularities in perturbation theory and sending $n\to
  1$, one finds that (\ref{rft}) is equivalent to the Parisi-Sourlas action \cite{PAS1}
\begin{widetext}
\begin{eqnarray} 
{\mathcal L}=\int dx\; \frac 1 2 (m_2+m_3) \omega^2 + \omega\left( 
-\Delta \phi + m_1\phi+3 g\phi^2
\right)+\frac 1 2\sum_{a,b}\chi_{ab} \left( 
-\Delta + m_1+6 g\phi
\right)\chi_{ab}\, ,
\label{fin}
\end{eqnarray} 
\end{widetext}
where the matrix of fields $\chi_{ab}(x)$ has $({n}-1)({n}-2)/2-2$
independent components. These fields enter quadratically in (\ref{fin}). 
Explicit integration over them gives rise to a determinant to the power that tends to $-2$ for $n\to 1$ and is 
equivalent to the one generated by fermion fields.
The cubic vertex has a coupling constant 
$g=\omega_1-\omega_2$, it has scaling dimension $\frac D 4-2$ and
becomes relevant below dimension 8.

In different systems the various parameters appearing in the action
are function of the physical control parameters e.g. temperature and
pinning field amplitude, or fraction of blocked particles. At the
critical point one has $\omega_1=\omega_2$, the cubic coupling constant $g$ 
is equal to zero and sub-leading singular vertexes need to be considered in the
Landau free-energy expansion. We should therefore consider quartic
terms and sub-leading cubic ones.  
Dimensional analysis, through repeated use of (\ref{trasf}), shows that
among all possible quartic replica symmetric invariants, the most
singular ones are\footnote{We thank T. Rizzo and L. Leuzzi for pointing us a
contribution previously neglected.}
\begin{eqnarray}
u_1{\rm Tr\;}\phi^4+ u_2\sum_{ab}\phi_{ab}^4+u_3 \sum_{abc} 
\phi_{ab}^2 \phi_{ac}\phi_{cb}
\end{eqnarray} 
that after the change of basis the leading terms (lowest scaling dimension)  
give rise to the RFIM terms
\begin{eqnarray}
(u_1+u_2-u_3)\left[\omega \phi^3+2  \phi^2\sum_{ab}\chi_{ab}^2\right]. 
\end{eqnarray} 
The scaling dimension of this vertex is $\frac D 2 -3$ which thus becomes
relevant below 6 dimension. In addition one needs in principle to consider 
the cubic
vertexes that do not vanish for $\omega_1=\omega_2$. Among these, 
the ones of higher scaling dimension are ${\rm Tr \;}\chi^3$
and $\sum_{ab}\chi_{ab}^3$, which for $\omega_1=\omega_2$ give rise to 
the term
$\omega_1\left({\rm Tr \;}\chi^3+\sum_{ab}\chi_{ab}^3\right)$. This term
does not have evident reasons to vanish. Its superficial scaling 
dimension (when multiplied by $d^Dx$) is $\frac D 4- \frac 3 2$ and 
in absence of additional cancellations is would also
become relevant below 6 dimensions. The relevance of these vertexes
would spoil the equivalence with the RFIM at the critical point. 

In order to investigate the possible relevance of these terms in the
action we could in principle analyze their contribution to the
correlation functions of the various fields 
in perturbation theory. To do this one needs to
keep into account the constraints (\ref{constraints}) which give rise
to some cumbersome combinatorics.

We have preferred therefore to turn back to the original replica field
theory (\ref{rft}) and compute the exact contribution of the cubic
vertexes to the self-energy and to the correlation functions to the one loop
level. The diagrams contributing to the self-energy of a generic cubic 
theory are depicted in fig. 2, they are the diagrams irreducible 
against cuts that separate the ending points \cite{tadpole}. 



\begin{figure}
\includegraphics[width=0.35\columnwidth]{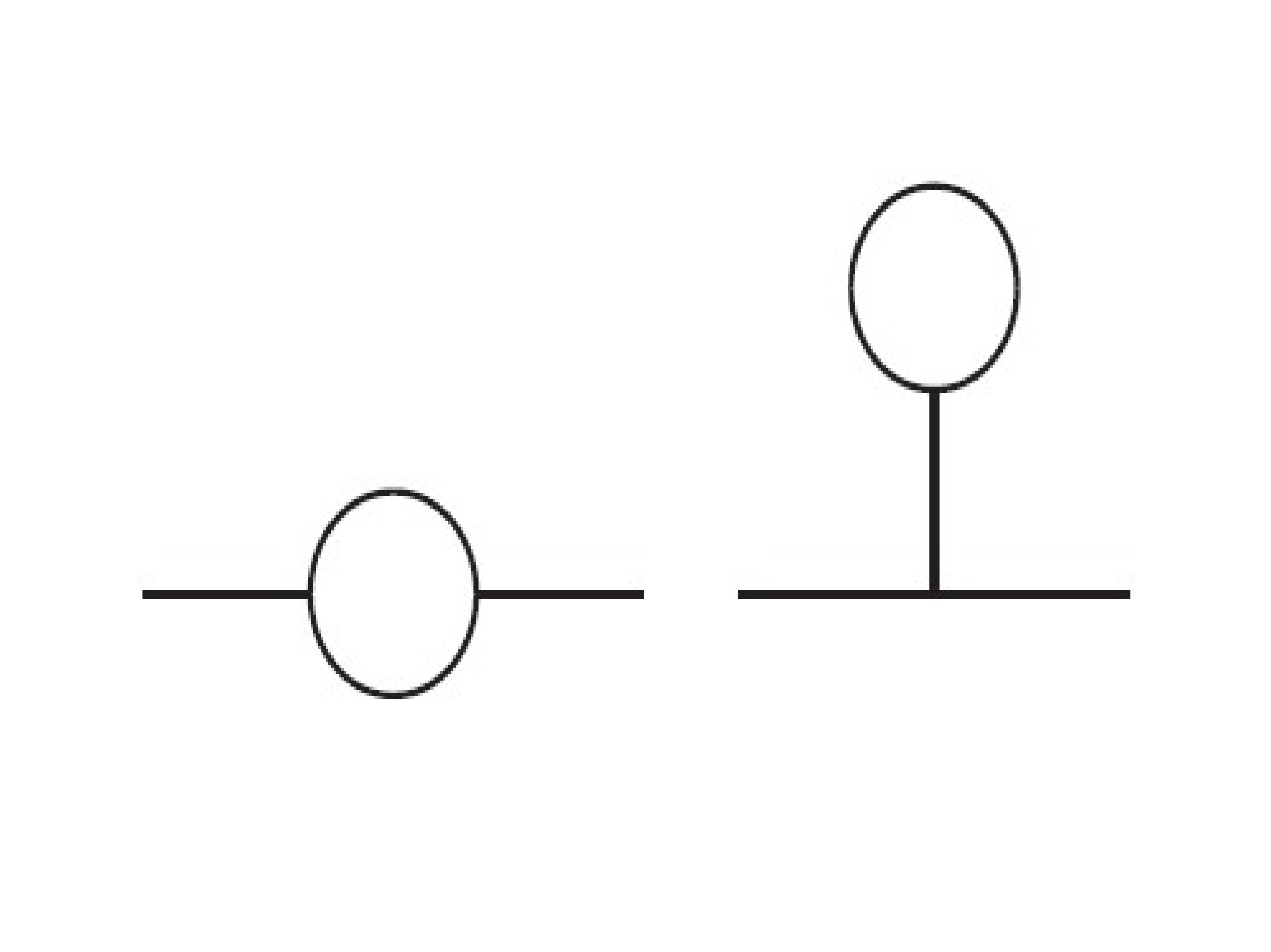}
\caption[0]{One-loop self-energy diagrams of a generic cubic theory.}
\label{fey1}
\end{figure}

%
The structure of the bare propagator of the theory in momentum space 
induced by replica symmetry is 
\begin{widetext}
\begin{eqnarray}
G_{ab;cd}(k)= (1-\delta_{ab})(1-\delta_{cd})\left[
g_1(k) \frac{\delta_{ac}\delta_{bd}+\delta_{ad}\delta_{bc}}{2}+g_2(k) 
\frac{\delta_{ac}+\delta_{bd}+\delta_{ad}+\delta_{bc}}{4}+g_3(k)
\right]
\label{bareprop}
\end{eqnarray}
\end{widetext}
where the three coefficients $g_i(k)$, $i=1,2,3$ can be expressed as
function of the mass parameters $m_i$ appearing in the action (\ref{rft}), in
the limit $n\to 1$:
\begin{eqnarray}
& &g_1(k)=\frac{2}{m_1 +k^2} \\
& &g_2(k)= \frac{4m_2}{(m_1 +k^2)(-2(m_1 +k^2)+m_2)}
\\
& &g_3(k)=\frac{2(-2m_3(m_1 +k^2)+m_2(m_2+m_3))}{(m_1 +k^2)^2(2(m_1 +k^2)-m_2)}.
\end{eqnarray}
Notice that while $g_1$ and $g_2$ have a simple pole behavior for small $k$ and $m_1$, $g_3$ has a double pole. 

Given the structure of the propagator and the vertexes, the direct manual
evaluation of the self-energy diagrams to include the sub-leading order is rather awkward. The formal expression of the diagrams of 
fig. 2 is given in the appendix. Fortunately, the computation of
the diagrams, which basically involves sums of constants and delta
functions over replica indexes and momentum integration, can be fully
automatized. We have achieved this goal using the algebraic
manipulator Mathematica, defining functions that implement the
Kronecker delta on the one hand and the sum of constants and delta
functions on the other.
In order to compute the contribution to the correlation function one
needs to further multiply the resulting
expression to the right and to the left by
the bare propagator, which can also be done by automatic means.

The result for the corrections to the 
correlation function can be cast in a form similar to the one of the 
bare propagator
(\ref{bareprop}), parametrized by three numbers $r_1, \;r_2\; r_3$
analogous to the $g$'s of (\ref{bareprop}). The resulting expressions,
available upon request, are rather lengthy and we do not reproduce them
here.  For $\omega_1\ne \omega_2$ the leading singularity behaves as
$m_1^{D/2-6}$, with corrective terms of order $m_1^{D/2-5}$ and
higher.  Comparing the leading contribution with the behavior of the
bare propagator $m_1^{-2}$ one sees that, as already found with the
$\phi^3$-RFIM mapping of \cite{gangof4}, non Gaussian fluctuations
become important below dimension 8. For  $\omega_1=\omega_2$ 
the leading terms vanish. Remarkably the computation shows that 
at the same time the
corrective terms of order $m_1^{D/2-5}$, also vanish. In fact these
contributions are proportional to $\omega_1-\omega_2$. The first non vanishing
contributions are of order $m_1^{D/2-4}$. These are irrelevant down to
dimension 4, as opposed to the quartic RFIM contribution which
become relevant in dimension 6. 

This result proves the $\phi^4$-RFIM mapping to the one loop order in
perturbation theory. Though we believe the result to be valid to all
orders the general mechanism for cancellation of the contribution of the
cubic vertexes remains to be found.
It would be tempting to think that for $n\to 1$ the fields
$\chi_{ab}(x)$ in (\ref{trasf}) verify fermionic algebraical relations
that go beyond the simple identification of the replica determinant
with the fermionic one that was used in \cite{gangof4}.

\section{Simulations}

In order to check whether the universality class of the terminal critical point is that of a $\phi^4$-RFIM we have simulated the 3-spin model on a random regular graph with fixed degree $z=8$. The Hamiltonian of the model is
\[
\mathcal{H} = -\sum_{\mu=1}^{zN/3} J_\mu S_{i_1^\mu} S_{i_2^\mu} S_{i_3^\mu}\;,
\]
where $S_i=\pm1$ are $N$ Ising variables and the $zN$ indexes $\{i_1^\mu,i_2^\mu,i_3^\mu\}$ are randomly chosen such that each variable appears exactly $z$ times and each interaction involves 3 different variables. The couplings are independent random binary variables, generated according to $\mathbb{P}(J_\mu=1)=r$ and $\mathbb{P}(J_\mu=-1)=1-r$.
In the thermodynamical limit, the model has been solved with the cavity method \cite{MRT04} and presents a dynamical phase transition at a temperature $T_d=1.3420(5)$ \cite{gangof4}, followed by Kauzmann ideal glass transition at a lower temperature $T_K$.
As long as $T>T_K$, the thermodynamical and dynamical properties of the model do not depend on the choice of the coupling bias $r$, because almost any coupling configuration has the same free energy, thanks to the fact that the annealed average equals the quenched one ($\log \mathbb{E}_J Z_J = \mathbb{E}_J \log Z_J$). In particular we can choose the couplings according to the Nishimori prescription \cite{Nishi} that corresponds to $r=(1+\tanh\beta)/2$. This choice has the great advantage that the all-spin-up configuration $S_i=1$ is an equilibrium configuration and we can use it as the starting configuration for the study of the equilibrium dynamics and also as the initial configuration for a Monte Carlo simulation, which does not require thermalization \cite{KZ}.

\subsection{Phase diagram with frozen variables}

We are interested in studying the 3-spin model defined above, when a fraction $\theta$ of variables are frozen to an equilibrium configuration (the all-spin-up configuration in the present case). The phase diagram in the $(\theta,T)$ plane resembles closely the one derived in \cite{MRS,RTS} for the 3-XORSAT model~\footnote{The 3-XORSAT model is nothing but the 3-spin model on a random graph of mean degree $3\alpha$ at $T=0$. Phase transitions takes place varying $\alpha$, which plays the same role of the temperature in thermal models.} in the $(\theta,\alpha)$ plane (see Fig.3 in \cite{RTS}).
It is also very similar to the one derived in \cite{BC} for the spherical 3-spin model.

\begin{figure}
\includegraphics[width=0.7\columnwidth]{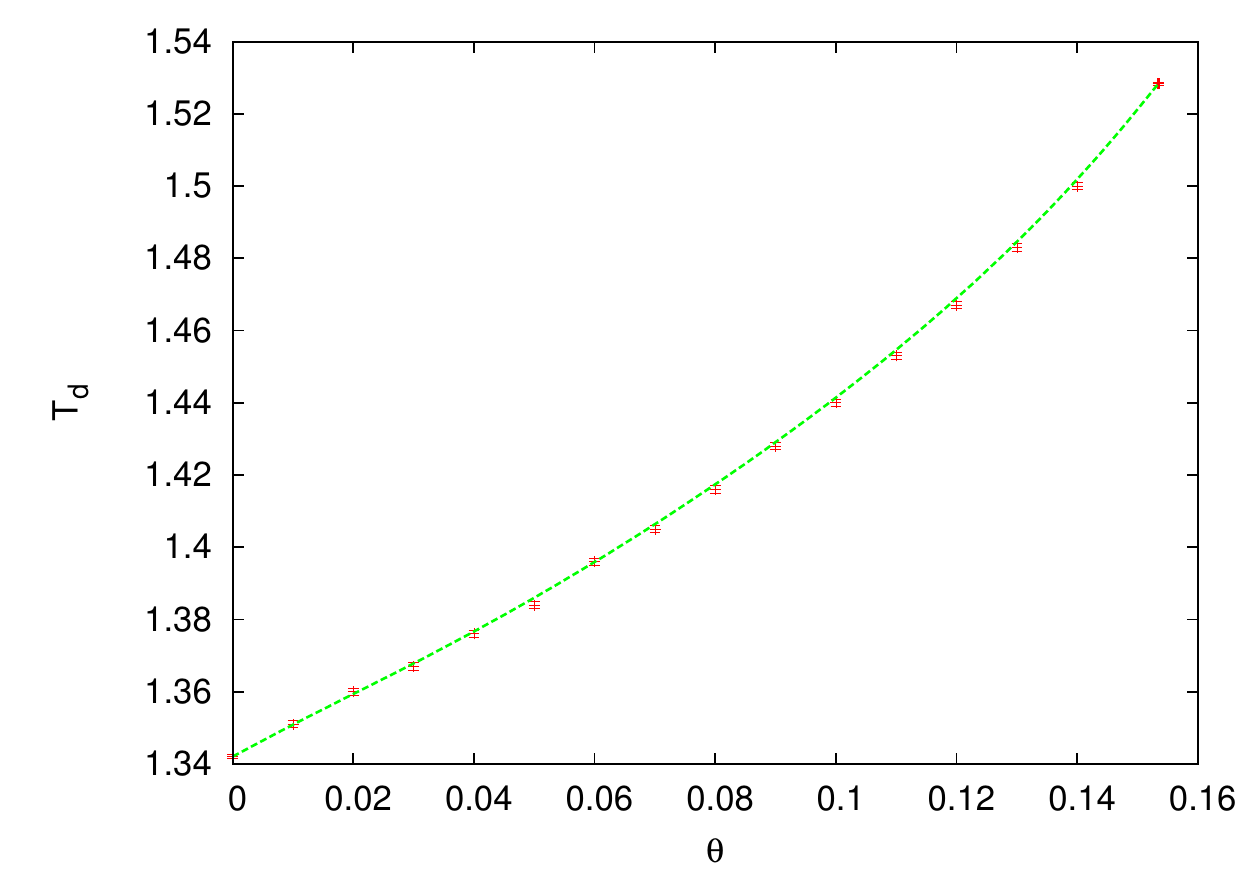}
\caption{The dynamical critical line as a function of the fraction $\theta$ of frozen variables in the 3-spin model on a random regular graph with a fixed degree $z=8$. Critical temperatures are shown with error; the curve is just a guide for the eyes, not a fit.}
\label{Td}
\end{figure}

In Fig.~\ref{Td} we show the numerical data for the dynamical critical temperatures as a function of $\theta$ obtained by the cavity method. All the critical points shown (but the terminal point that we discuss in the following) as been obtained by looking at the overlap with the reference equilibrium configuration ($S_i=1$): the dynamical critical temperature corresponds to the negative jump in this overlap when the temperature is slowly increased.

A much more careful discussion requires the determination of the terminal critical point, that we estimate to be located in $\theta_c=0.1534(2)$ and $T_c=1.5277(5)$.
Indeed at the terminal critical point the phase transition is no longer of first order and there is no jump in the overlap with the reference configuration to be exploited (even for $\theta<\theta_c$ but close to the terminal point the jump is so small that is not useful to estimate the transition location). On the contrary, for $\theta>\theta_c$ the model has no phase transition at all and one can at most look at the precursors of the continuous phase transition taking place at the terminal critical point.

The problem of locating the terminal critical point with high accuracy is delicate and we have followed a method based on solving the cavity equations in population, while looking at the instability parameter $\lambda$. More precisely, we run the Belief Propagation (BP) algorithm for determining the fixed point distribution of cavity magnetizations $P(m)$ that solves the equation
\[
P(m) = \theta\:\delta(m-1) + (1-\theta)\:\mathbb{E}_J \int \prod_{i=1}^{z-1} P(m^i_1) dm^i_1 P(m^i_2) dm^i_2\;\:
\delta\left(m-J\frac{\prod_i (1+t_\beta m^i_1m^i_2) - \prod_i (1-t_\beta m^i_1m^i_2)}{\prod_i (1+t_\beta m^i_1m^i_2) + \prod_i (1-t_\beta m^i_1m^i_2)}\right)\;,
\]
where $t_\beta=\tanh(\beta)$.
Then we have computed the local stability of the fixed point distribution by adding a small perturbation to each element $m$ of the population and checking whether such a perturbation grows or decreases under BP iterations: the stability parameter $\lambda$ is defined such that the perturbation goes like $e^{\lambda t}$ for large times $t$. So the fixed point is stable only if $\lambda < 0$. In a discontinuous transition the condition $\lambda=0$ identifies the spinodal points where the two states becomes locally unstable; while in a continuous transition the condition $\lambda=0$ marks the unique critical point.

\begin{figure}
\includegraphics[width=0.7\columnwidth]{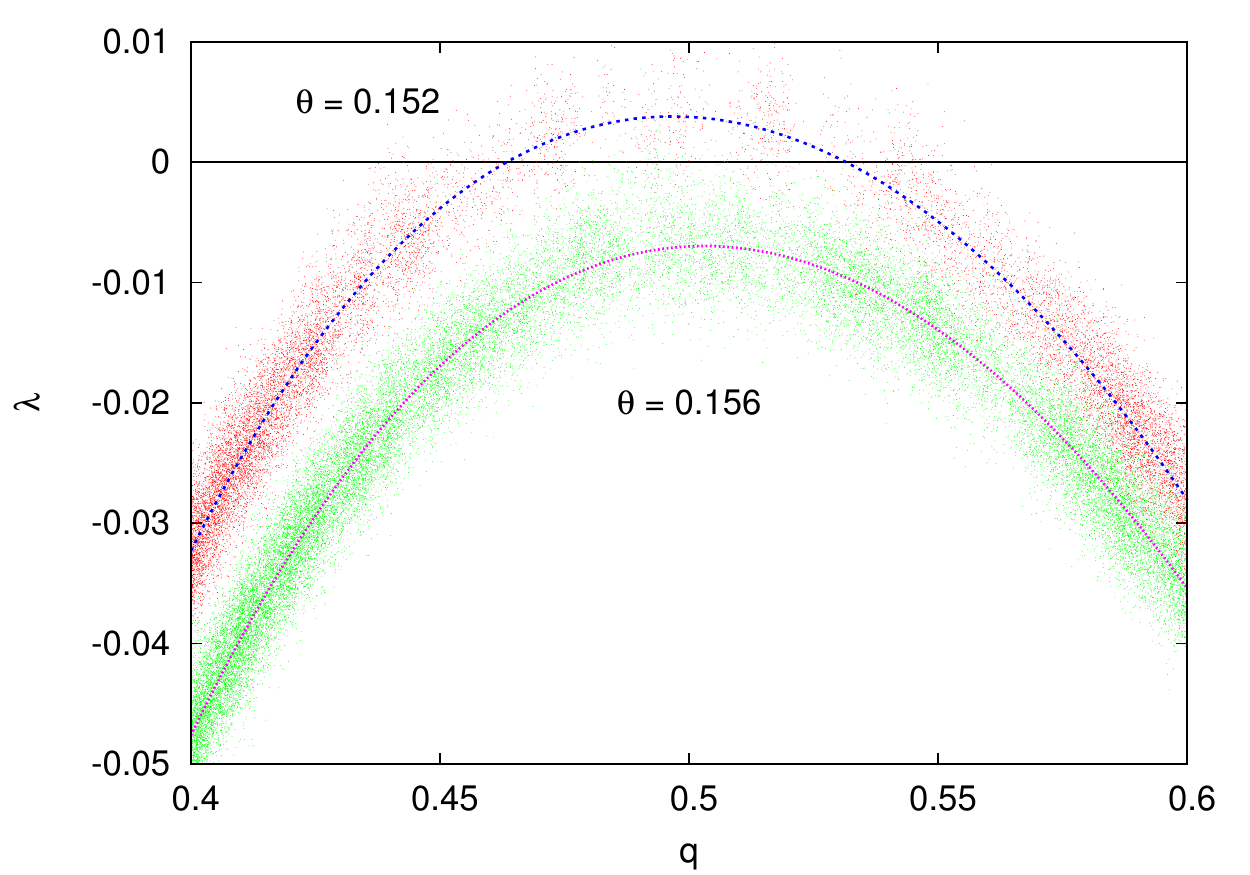}
\caption{The stability parameter $\lambda$ as a function of the overlap $q$ with the reference configuration for two values of the fraction of frozen variables $\theta=0.512$ and $\theta=0.156$. Data points are from fixed point distributions measured at several different temperatures around the critical one. Curves are the best fitting quartic polynomials. Data at $\theta=0.152$ show a discontinuous transition, while data at $\theta=0.156$ show the absence of any critical point. In this plot, the terminal critical point $\theta_c$ would correspond to a maximum $\lambda$ value equal to 0.}
\label{lambda}
\end{figure}

In Fig.~\ref{lambda} we show the stability parameter $\lambda$ as a function of the overlap $q$ with the equilibrium reference configuration. For each value of $\theta$, the plot contains few tens of thousands of points measured at different temperatures around the critical one (both above and below the critical temperature). The noise in the data is due to the stochastic nature of the BP algorithm and to fluctuations related to the finiteness of the population (we have used populations of sizes $10^6$).
It is worth stressing that $\lambda$ mainly depends on $q$, while its dependence on the temperature is rather weak: indeed Fig.~\ref{lambda} shows data measured at different temperatures that lay on the same curve. The interpolating curves shown in Fig.~\ref{lambda} are quartic polynomials.

For $\theta=0.152$ the model undergoes a discontinuous phase transition: the two $q$ values for which $\lambda=0$ bracket an interval where no state can exist in the thermodynamical limit (since it would be locally unstable); by varying the temperature, the thermodynamic overlap has a jump not smaller than the size of this interval. Moreover the larger overlap where $\lambda=0$ is the plateau value at the dynamical critical temperature $T_d$.
For $\theta=0.156$ the stability parameter $\lambda$ never becomes positive and consequently the model has a unique paramagnetic phase and no phase transition.
The $\theta_c$ value of the terminal critical point is clearly in between these two $\theta$ values and for $\theta=\theta_c$ the $\lambda(q)$ function must have a maximum of height $\lambda=0$.
The quartic interpolations shown in Fig.~\ref{lambda} allow us to obtain reliable estimate of position and height of the maximum of the function $\lambda(q)$ for each $\theta$ value. Fitting these maxima we then arrive at the following estimates for the terminal critical point: $\theta_c=0.1534(2)$ and $q_c=0.499(1)$.

\begin{figure}
\includegraphics[width=0.7\columnwidth]{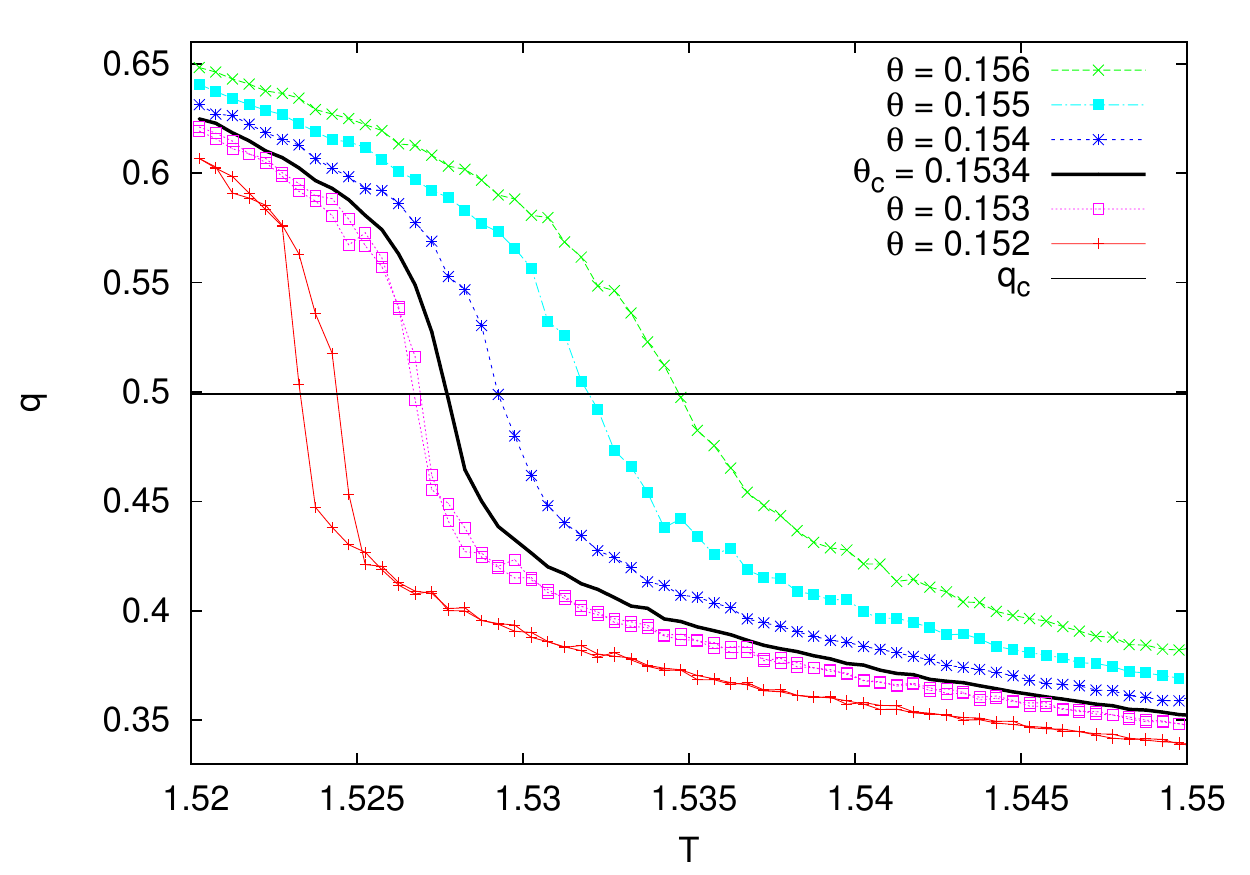}
\caption{Average value for the overlap with the equilibrium reference configuration as a function of the temperature for several $\theta$ values. The lowest dataset shows a clear hysteresis cycle at the first order phase transition. The data plotted with a black line have been measured at the critical $\theta_c$. The horizontal line is the critical value for the overlap.}
\label{mediaQ}
\end{figure}

A reliable way to estimate the temperature of the terminal critical point is by plotting the average overlap as a function of the temperature for $\theta$ values at and around $\theta_c$ (see Fig.~\ref{mediaQ}). In this way we can estimate $T_c$ as the temperature at which the average overlap measured at $\theta_c$ reaches the critical value $q_c$. The uncertainty on this value comes both form the statistical error in this fitting procedure and from the propagation of the uncertainty on the value $\theta_c$. In Fig.~\ref{mediaQ} we plot data for several $\theta$ values such that one can appreciate how much the average overlap changes by varying $\theta$. The final estimate for the temperature of the terminal critical point is $T_c=1.5277(5)$. The error is mainly given by the uncertainty on the value of $\theta_c$.

\subsection{Fluctuations at the terminal critical point}

Once we have obtained a reliable estimate for the location of the terminal critical point, $\theta_c=0.1534(2)$ and $T_c=1.5277(5)$, we have run extensive Monte Carlo simulations for those critical parameters. We have simulated sizes up to $N=3\cdot 10^4$. For each size, but the largest, we have simulated $10^4$ different samples (for the largest size only 2500 samples were used). Thermalization is not an issue, given that the all-spin-up configuration is an equilibrium configuration by construction.

As already discussed in detail in \cite{gangof4}, there are 3 different sources of randomness in this model: the coupling configuration, the starting equilibrium spin configuration and the thermal noise. Thanks to the fact the annealed approximation is exact in this model above $T_K$, in the thermodynamical limit every sample behaves exactly the same, when the average over starting configurations and thermal noise is performed.
However, we do not take the average over many initial spin configurations, and we only use $S_i=1$ as the starting configuration to avoid thermalization. So, the average over the many coupling configurations we consider does actually correspond to the average over the initial spin configurations: indeed we could gauge transform each sample in order to have roughly the same couplings and the initial spin configuration would change from sample to sample.

So, we are left we only 2 sources of randomness: heterogeneities in the initial configuration (\emph{het}) and thermal noise (\emph{th}). Using the angular brackets for the thermal average and the square brackets for the average over the initial spin configurations, two different susceptibilities can be defined as follows
\begin{eqnarray}
\chi_{th} &=& N \left[\langle q^2 \rangle - \langle q \rangle^2 \right]\;,\\
\chi_{het} &=& N \left(\left[\langle q \rangle^2\right] - [\langle q \rangle]^2\right)\;,
\end{eqnarray}
where the former measures thermal fluctuations within the same sample (averaged over the samples), while the latter quantifies sample to sample fluctuations in the mean value of $q$ (which is as usual the overlap with the equilibrium reference configuration). The total susceptibility is the sum of the two: $\chi_{tot} = N\left(\left[\langle q^2 \rangle\right] - [\langle q \rangle]^2\right) = \chi_{th} + \chi_{het}$.

\begin{figure}
\includegraphics[width=0.7\columnwidth]{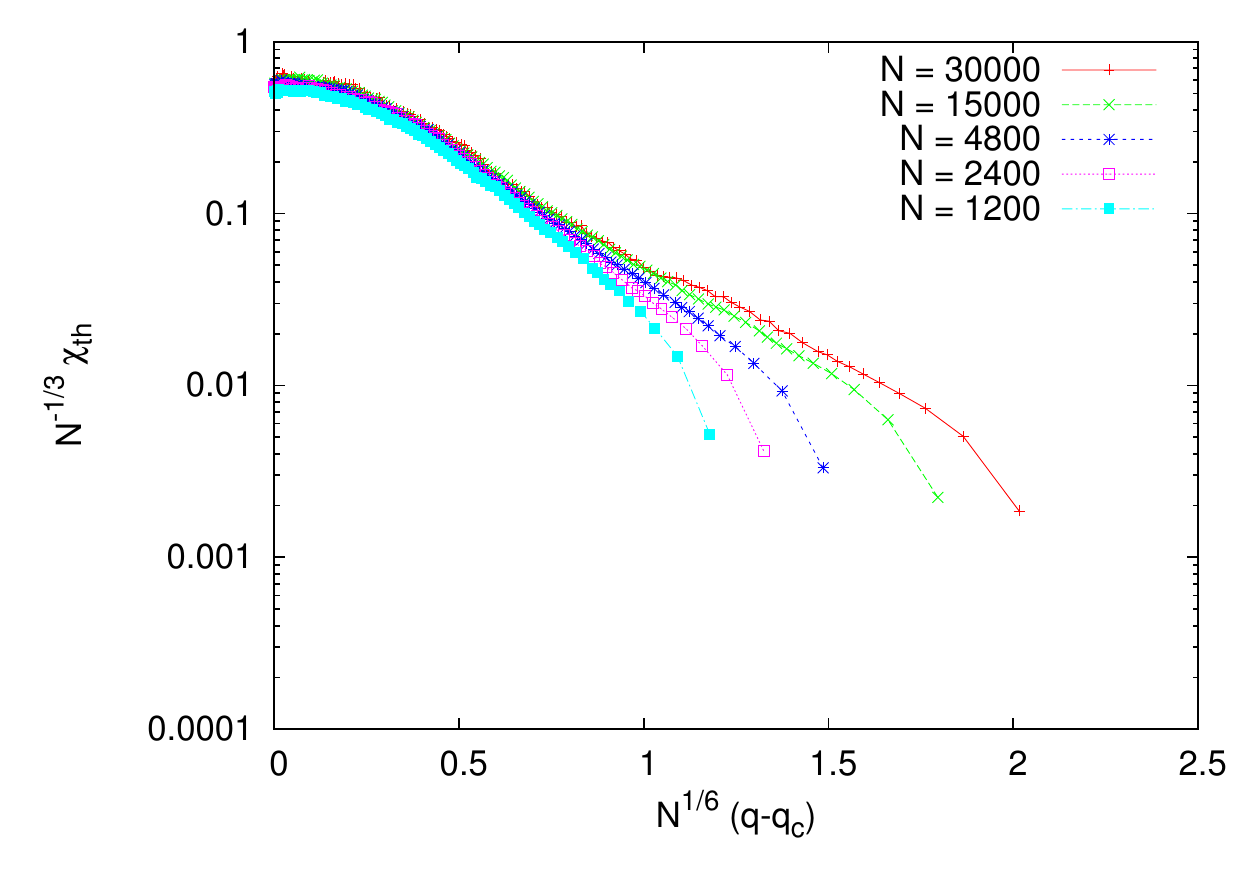}
\includegraphics[width=0.7\columnwidth]{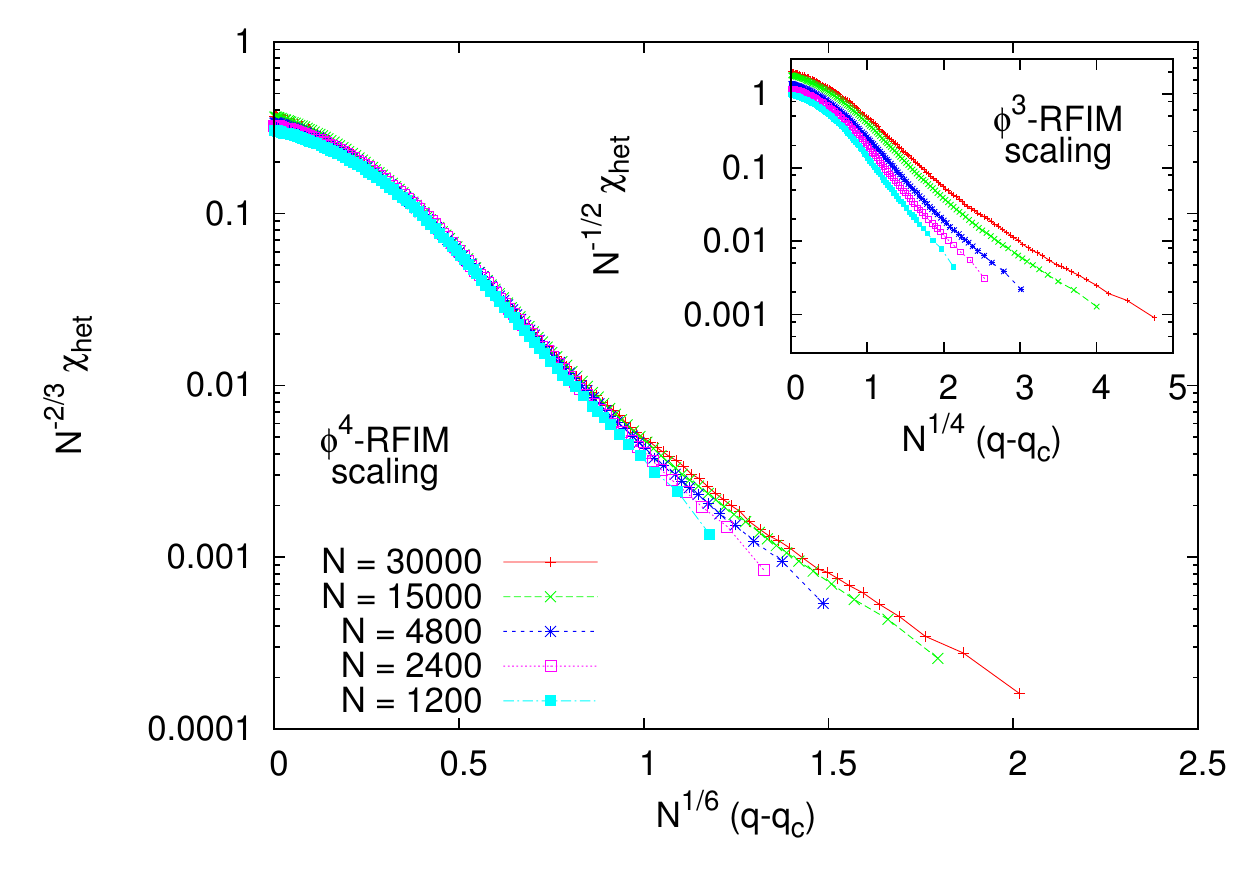}
\caption{Fluctuations over the thermal noise $\chi_{th}$ (above) and over the initial heterogeneities $\chi_{het}$ (below) rescaled according to the exponents predicted by a $\phi^4$-RFIM theory, for comparison in the inset we plot 
the same quantity using the scaling appropriate for the $\phi^3$-RFIM. 
The data have been measured at the terminal critical point of the diluted 3-spin model by Monte Carlo simulations.}
\label{scaled}
\end{figure}

A dimensional analysis of action of the $\phi^4$-RFIM leads straightforwardly to the following scaling relations for the susceptibilities
\begin{eqnarray}
\chi_{th} &=& N^{1/3}\:\widehat\chi_{th}\left(N^{1/6}(q-q_c)\right)\;,\\
\chi_{het} &=& N^{2/3}\:\widehat\chi_{het}\left(N^{1/6}(q-q_c)\right)\;,
\end{eqnarray}
In Fig.~\ref{scaled} we show the data collapse using the theoretically predicted exponents. We can see that the scaling for the largest fluctuations, $\chi_{het}$, is very well verified, while some corrections to the scaling are still present in the data for the thermal fluctuations at the sizes we have simulated.

\section{Summary}

Summarizing, we have found that Replica Field Theory of glasses
predicts that fluctuations close to glassy critical points can be
mapped into the Random Field Ising Model. This observation calls for
the search of glassy critical points in liquid systems. In simulations
one can use the method of pinning field or blocking a fraction of the
particles. In experiments critical points could be found in liquids
confined by porous media.

{\bf Acknowledgments}

We thank G.Biroli, C.Cammarota, L. Leuzzi, T. Rizzo and P. Urbani 
for discussions. 
SF acknowledges the physics department of Rome university ``Sapienza'' for
hospitality. F. R.-T. acknowledges the LPTMS, Universit\'e Paris-Sud 11
for hospitality. The European Research Council has provided financial
support through ERC grant agreement no. 247328.

\section{Appendix}
This appendix is devoted to the explanation of some technical details of 
the evaluation of the one loop corrections to the propagator. 

The diagrams that we need to evaluate are conceptually simple, but practically 
complicated by the presence of the replica indexes in the propagator and the vertexes that we write as: 
\begin{widetext}
\begin{eqnarray}
G_{ab;cd}(k)= (1-\delta_{ab})(1-\delta_{cd})\left[
g_1(k) \frac{\delta_{ac}\delta_{bd}+\delta_{ad}\delta_{bc}}{2}+g_2(k) 
\frac{\delta_{ac}+\delta_{bd}+\delta_{ad}+\delta_{bc}}{4}+g_3(k)
\right]
\label{barepropapp}
\end{eqnarray}
\end{widetext}
\begin{widetext}
\begin{eqnarray}
\Gamma_{ab;cd;ef}= {\bf S} [\omega_1 \delta_{ac} \delta_{ae} \delta_{bd} \delta_{be}+\omega_2 \delta_{bc} \delta_{de} \delta_{fa}]
\label{vertex}
\end{eqnarray}
\end{widetext}
where ${\bf S}$ denotes symmetrization against exchange of indexes in
each couple $(ab)$ $(cd)$ and $(ef)$ and exchange of the couples among
themselves. The first self-energy diagram reads
\begin{widetext}
\begin{eqnarray}
\Sigma_{ab;cd}^{(1)}(k)=\int dq \; & &{\bf S} \left(
\omega_1^2
{ \sum}_{e,f}  G_{ae;cf}(q)G_{be;df}(k+q)+
\right. 
\\
& & \omega_1 \omega_2 {\sum}_{e}[G_{ae;cd}(q)G_{be;cd}(k+q)+G_{ab;ce}(q)G_{ab;de}(k+q)]+\\
& &\left. 
\omega_2^2 G_{ab;cb}(q)G_{ab;cb}(k+q)
\right)\\
\label{self1}
\end{eqnarray}
\end{widetext}
the second one is 
\begin{widetext}
\begin{eqnarray}
\Sigma_{ab;cd}^{(2)}(k)=
& {\bf S}\int dq\;\left( \omega_1^2
\sum_{e,f,g} G_{ef;fg}(q)G_{eg;ac}(0)\delta_{bd}
+\omega_2^2 \sum_{e,f} G_{ef;ef}(q)G_{ef;ab}(0)\delta_{ac}\delta_{bd}+
\right.
\nonumber\\
&\left.\omega_1\omega_2 
\sum_{e,f,g} G_{ef;fg}(q)G_{eg;ab}(0)\delta_{ac}\delta_{bd}+
\omega_1\omega_2 \sum_{ef}
G_{ef;ef}(q)G_{ef;ac}(0)\delta_{bd}\right).
\label{self2}
\end{eqnarray}
\end{widetext}
In order to compute the correction to the propagator the one loop self-energy 
$\Sigma^{1L}_{ab;cd}(k)$ should be multiplied to the right and to the left by the 
bare propagator. 
\begin{widetext}
\begin{eqnarray}
  \delta G_{ab;cd}(k)=\sum_{a',b',c',d'} G_{ab;a'b'}(k)\Sigma_{a'b';c'd'}^{1L}(k)
G_{c'd';cd}(k).
\end{eqnarray}
\end{widetext}
The whole calculation involves sums of constants and delta functions
over replica indexes. In order to avoid mistakes we have automatized the calculation through the use of the software Mathematica. We reproduce here the commented worksheet
used in the calculation.

\newpage
\mbox{Mathematica worksheet}


\section*{Supplementary material}

Mathematica worksheet

```
In[1]:=  (* Definition of Kronecker Delta Kd[a-b] *)

In[2]:=  Unprotect[Times] ; Unprotect[Power]

Out[2]=  {Power}

In[3]:=  Kd[0] := 1
         Kd[-a_] := Kd[a]
         Kd[a_] Kd[a_] := Kd[a]
         Kd[a_]^1 := Kd[a]
         Kd[a_]^n_ := Kd[a] /; n > 1
         Kd[x_] := 0 /; x ≠ 0

In[9]:=  (* End Definition of Kronecker Delta Kd[a-b] *)

In[10]:=
         (* Definition of sum *)

In[11]:= sum[Kd[a_ - b_] x_, a_] := x /. a → b
         sum[Kd[a_ - b_] x_, b_] := x /. b → a
         sum[Kd[a_ - b_], b_] := 1
         sum[Kd[a_ - b_], a_] := 1
         sum[Kd[a_ - 1], a_] := 1
         sum[Kd[a_ - 2], a_] := 1
         sum[Kd[-1 + a_], a_] := 1
         sum[Kd[-2 + a_], a_] := 1
         sum[y_ x_, a_] := y sum[x, a] /; FreeQ[y, a]
         sum[y_ + x_, a_] := sum[x, a] + sum[y, a]
         sum[y_, a_] := y n /; FreeQ[y, a]

In[22]:= (* End Definition of sum *)


In[23]:= (* Propagator : to be considered a matrix in the couples of indexes (ab) and (cd) *)

In[24]:= G[a_, b_, c_, d_] :=
           Expand[(1 - Kd[a - b]) (1 - Kd[c - d]) (g1/2 (Kd[a - c] Kd[b - d] + Kd[a - d] Kd[b - c]) +
             g2/4 (Kd[a - c] + Kd[b - d] + Kd[a - d] + Kd[b - c]) + g3)];

In[25]:=
         (* Increase the recursion limit *)

In[26]:= $RecursionLimit = 7000

Out[26]= 7000

In[27]:= (* Matrix with the same structure of the propagator and different parameters *)

In[28]:= M[x_, y_, c_, d_] = G[x, y, c, d] /. {g1 → m1, g2 → m2, g3 → m3};

         (* Matrix product *)

In[29]:= Uno[a_, b_, c_, d_] := Evaluate[sum[sum[Expand[G[a, b, x, y] M[x, y, c, d]], x], y]]

         (* Notice that the product has the same structure of G and M *)

In[30]:= (* Find g1,g2,g3 such that G is the inverse matrix of M *)

In[31]:= inv = FullSimplify[
           Solve[{Uno[1, 2, 1, 2] == 1, Uno[1, 2, 1, 3] == 0, Uno[1, 2, 3, 4] == 0}, {g1, g2, g3}]] /. n → 1
```

$$\text{Out[31]= } \left\{\left\{g1 \to \frac{2}{m1},\ g2 \to \frac{4\ m2}{m1\ (-2\ m1 + m2)},\ g3 \to \frac{2\ (-2\ m1\ m3 + m2\ (m2 + m3))}{m1^2\ (2\ m1 - m2)}\right\}\right\}$$

```
In[32]:= (* Introduce the momentum dependence ; k is adimensional *)
```

```
In[33]:= invk = inv /. m1 → m1 (1 + k^2)
```

$$\text{Out[33]}= \left\{\left\{g1 \to \frac{2}{\left(1+k^2\right) m1},\ g2 \to \frac{4\, m2}{\left(1+k^2\right) m1 \left(-2\left(1+k^2\right) m1 + m2\right)},\right.\right.$$
$$\left.\left. g3 \to \frac{2\left(-2\left(1+k^2\right) m1\, m3 + m2\, (m2+m3)\right)}{\left(1+k^2\right)^2 m1^2 \left(2\left(1+k^2\right) m1 - m2\right)}\right\}\right\}$$

```
In[34]:=

In[35]:= (* Representation in term of matrix
          elements : The replica symmetrix matrices have elements of three types:
               1) the couples of indexes coincide (ab)=(cd) 2) there is one
          index in common e.g. a=c but b ≠ d 3) the indexes are all different. Calling x1,
   x2, x3 these elements one can express g1,g2,g3 as a function of the x's *)

In[36]:= ss = Solve[{x1, x2, x3} == {G[1, 2, 1, 2], G[1, 2, 1, 3], G[1, 2, 3, 4]}, {g1, g2, g3}][[1]]

Out[36]= {g1 → 2 (x1 - 2 x2 + x3), g2 → 4 (x2 - x3), g3 → x3}

In[37]:= (* Same for the m's *)

In[38]:= ss1 = ss /. {g1 → m1, g2 → m2, g3 → m3, x1 → y1, x2 → y2, x3 → y3}

Out[38]= {m1 → 2 (y1 - 2 y2 + y3), m2 → 4 (y2 - y3), m3 → y3}

In[39]:= (* ... and the elements of the matrix product, for n→1 *)
        uu = Evaluate[Simplify[{Uno[1, 2, 1, 2], Uno[1, 2, 1, 3], Uno[1, 2, 3, 4]} /. ss /. ss1 /. n → 1]]

Out[39]= {2 (x1 y1 - 2 x2 y2 + x3 y3), 2 (x1 y2 - 2 x3 y2 + x2 (y1 - y2 - 2 y3) + 3 x3 y3),
          2 (4 x2 y2 + x1 y3 - 6 x2 y3 + x3 (y1 - 6 y2 + 6 y3))}

In[40]:= (* Elements of product : expression above written explicitly *)

In[41]:= prod[x1_, x2_, x3_, y1_, y2_, y3_] :=
         {2 (x1 y1 - 2 x2 y2 + x3 y3), 2 (x1 y2 - 2 x3 y2 + x2 (y1 - y2 - 2 y3) + 3 x3 y3),
          2 (4 x2 y2 + x1 y3 - 6 x2 y3 + x3 (y1 - 6 y2 + 6 y3))}

In[42]:= (* End Matrix Product *)

In[43]:=
        (* Computation of diagrams *)

In[44]:= (* First Self-energy diagram *)
        (* Building up the w1 - w1 sub-diagram : H11[a,b,c,d] *)

In[45]:= h = Expand[G[a, c, a1, c1] (G[b, c, b1, c1] /. {g1 → gg1, g2 → gg2, g3 → gg3})];
        (* two internal propagators attached to w1 - w1 vertexes ; gg1, gg2,
        gg3 will be useful later to take into account momentum dependence *)

In[46]:= G11[a_, b_, a1_, b1_] := Evaluate[(1 - Kd[a - b]) (1 - Kd[a1 - b1]) sum[sum[h, c], c1]]
        (* Sum over internal indexes *)

In[47]:= H11[a_, b_, a1_, b1_] :=
         (G11[a, b, a1, b1] + G11[b, a, a1, b1] + G11[a, b, b1, a1] + G11[b, a, b1, a1])/
           4 (* Symmetrization *)


        (* Building the w2 - w2 sub-diagram : H22[a,b,c,d] *)

In[48]:= G22[a_, b_, a1_, b1_] := G[a, b, a1, b1] (G[a, b, a1, b1] /. {g1 → gg1, g2 → gg2, g3 → gg3})
            (* two internal propagators attached to w2 - w2 vertexes *)

In[49]:= H22[a_, b_, a1_, b1_] := G22[a, b, a1, b1]


        (* Building the w1 - w2 sub-diagram  H12[a,b,c,d] *)

In[50]:= p = sum[Expand[G[a, c, a1, b1] (G[b, c, a1, b1] /. {g1 → gg1, g2 → gg2, g3 → gg3})], c];
        (* two internal propagators attached to w1 - w2 vertexes + sum of one internal index *)
```

```
In[51]:= G12[a_, b_, a1_, b1_] := Evaluate[p (1 - Kd[a - b])]
```

```
In[52]:= H12[a_, b_, a1_, b1_] := 1/2 (G12[a, b, a1, b1] + G12[b, a, a1, b1])  (* Symmetrization *)
```

```
In[53]:= Diag[a_, b_, c_, d_] := w1^2 H11[a, b, c, d] + 2 w1 w2 H12[a, b, c, d] +
      w2^2 H22[a, b, c, d]  (* Total self-energy diagram  (sum of the three) *)
```

```
In[54]:= invk1 = invk /. {g1 → gg1, g2 → gg2, g3 → gg3, k → q}
```

$$\text{Out[54]= } \left\{\left\{gg1 \to \frac{2}{m1\left(1+q^2\right)}, gg2 \to \frac{4\,m2}{m1\left(1+q^2\right)\left(m2 - 2\,m1\left(1+q^2\right)\right)}, \right.\right.$$
$$\left.\left. gg3 \to \frac{2\left(m2\left(m2+m3\right) - 2\,m1\,m3\left(1+q^2\right)\right)}{m1^2\left(1+q^2\right)^2\left(-m2 + 2\,m1\left(1+q^2\right)\right)}\right\}\right\}$$

```
In[55]:= (* Substitution of parameters of propagator with inverse of momentum dependent mass *)
```

```
In[56]:= ii = Join[invk[[1]], invk1[[1]]]
```

$$\text{Out[56]= } \left\{g1 \to \frac{2}{\left(1+k^2\right) m1}, g2 \to \frac{4\,m2}{\left(1+k^2\right) m1\left(-2\left(1+k^2\right) m1 + m2\right)}, \right.$$
$$g3 \to \frac{2\left(-2\left(1+k^2\right) m1\,m3 + m2\left(m2+m3\right)\right)}{\left(1+k^2\right)^2 m1^2 \left(2\left(1+k^2\right) m1 - m2\right)}, gg1 \to \frac{2}{m1\left(1+q^2\right)},$$
$$\left. gg2 \to \frac{4\,m2}{m1\left(1+q^2\right)\left(m2 - 2\,m1\left(1+q^2\right)\right)}, gg3 \to \frac{2\left(m2\left(m2+m3\right) - 2\,m1\,m3\left(1+q^2\right)\right)}{m1^2\left(1+q^2\right)^2\left(-m2 + 2\,m1\left(1+q^2\right)\right)}\right\}$$

```
In[57]:= Diag1[a_, b_, c_, d_] = Diag[a, b, c, d] /. ii;
```

```
In[58]:= (* Multiply external propagators at zero momentum to the self-energy *)
```

```
In[59]:= (* Correction to the propagator : An integration over q is needed,
     remembering that q is a-dimensional and the integral should be multiplied by m1^(D/2) *)
```

```
In[60]:= F = Simplify[prod[G[1, 2, 1, 2], G[1, 2, 1, 3], G[1, 2, 3, 4],
         Diag1[1, 2, 1, 2], Diag1[1, 2, 1, 3], Diag1[1, 2, 3, 4]] /. (inv) /. n → 1][[1]];
```

```
In[61]:= dG = Simplify[
       prod[G[1, 2, 1, 2], G[1, 2, 1, 3], G[1, 2, 3, 4], F[[1]], F[[2]], F[[3]]] /. (inv) /.
         n → 1 /. k → -q][[1]];
```

```
In[62]:= (* Series expansion of dG up to 1/m1^4 terms *)
```

```
In[63]:= final = Series[dG, {m1, 0, -5}]
```

$$\text{Out[63]= } \left\{\frac{16\,(m2+m3)^2\,(5+4\,q^2)\,(w1-w2)^2}{\left(1+q^2\right)^4 m1^6} + \frac{16\,(m2+m3)\,(2+q^2)\,(w1-w2)\,(5\,w1-w2)}{\left(1+q^2\right)^3 m1^5} + \frac{1}{O[m1]^4},\right.$$
$$\frac{16\,(m2+m3)^2\,(5+4\,q^2)\,(w1-w2)^2}{\left(1+q^2\right)^4 m1^6} + \frac{16\,(m2+m3)\,(w1-w2)\,(12\,w1 + 5\,q^2\,w1 - 4\,w2 - q^2\,w2)}{\left(1+q^2\right)^3 m1^5} +$$
$$\frac{1}{O[m1]^4}, \frac{16\,(m2+m3)^2\,(5+4\,q^2)\,(w1-w2)^2}{\left(1+q^2\right)^4 m1^6} +$$
$$\left.\frac{16\,(m2+m3)\,(w1-w2)\,(13\,w1 + 5\,q^2\,w1 - 5\,w2 - q^2\,w2)}{\left(1+q^2\right)^3 m1^5} + \frac{1}{O[m1]^4}\right\}$$

```
In[64]:= (* At the critical point w2=
     w1 both the terms of order m1^(-6) and the terms m1^(-5) vanish *)
```

In[65]:= ```
final_critical = Simplify[Series[
    Simplify[prod[G[1, 2, 1, 2], G[1, 2, 1, 3], G[1, 2, 3, 4], F[[1]], F[[2]], F[[3]]] /. inv /.
      n → 1 /. w2 → w1 /. k → q][[1]], {m1, 0, -4}]]
```

Out[65]= $\left\{ \dfrac{12\,w1^2}{\left(1+q^2\right)^2 m1^4} + \dfrac{1}{O[m1]^3},\ \dfrac{12\,w1^2}{\left(1+q^2\right)^2 m1^4} + \dfrac{1}{O[m1]^3},\ \dfrac{12\,w1^2}{\left(1+q^2\right)^2 m1^4} + \dfrac{1}{O[m1]^3} \right\}$

In[66]:= ```
(* The final result is that both the term 1/m1^6 and 1/m1^5 vanish for w1=
  w2 therefore the cubic vertexes can be neglected with respect to the quartic ones.
    Thee moment integration provides a factor m1^(D/2),
 so that the term is of order m1^(D/2-4),
 to be compared to the bare contribution m1^(-2),
 telling that the terms start to contribute in D=4. The quartic terms on the other
   hand are of the order m1^(D/2-5) which become relevant in dimension 6.  *)

(* The same is true for the second self-energy diagram *)
```

In[67]:= `(* Second Diagram : Ball *)`

In[68]:=

In[69]:= `(* Ball without external legs *)`

In[70]:= ```
ball[a_, b_] =
 Expand[(sum[G[c, a, c, b], c] w1 + G[a, b, a, b] w2) /. {g1 → gg1, g2 → gg2, g3 → gg3}]
(* First Ring *)
```

Out[70]= $-\dfrac{gg2\,w1}{2} - 2\,gg3\,w1 + \dfrac{gg2\,n\,w1}{4} + gg3\,n\,w1 + \dfrac{gg1\,w2}{2} + \dfrac{gg2\,w2}{2} + gg3\,w2 - \dfrac{1}{2}\,gg1\,w1\,Kd[a-b] +$
$gg3\,w1\,Kd[a-b] + \dfrac{1}{2}\,gg1\,n\,w1\,Kd[a-b] + \dfrac{1}{4}\,gg2\,n\,w1\,Kd[a-b] - \dfrac{1}{2}\,gg1\,w2\,Kd[a-b] -$
$\dfrac{1}{2}\,gg2\,w2\,Kd[a-b] - gg3\,w2\,Kd[a-b] + \dfrac{1}{4}\,gg2\,w2\,Kd[-a+b] - \dfrac{1}{4}\,gg2\,w2\,Kd[a-b]\,Kd[-a+b]$

In[71]:= `(* Attach a propagator to the ball *)`

In[72]:= `K[a_, b_] = Evaluate[sum[sum[Expand[ball[c, d] G[a, b, c, d]], c], d]];`

In[73]:=
`(* Attach an external propagator with a w1 vertex *)`

In[74]:= `J[a_, c_, a1_, b1_] = sum[Expand[K[a, b] G[c, b, a1, b1]], b];`

In[75]:= `(* Addition of external legs *)`

In[76]:= ```
F1 = w1 Simplify[prod[G[1, 2, 1, 2], G[1, 2, 1, 3],
     G[1, 2, 3, 4], J[1, 2, 1, 2], J[1, 2, 1, 3], J[1, 2, 3, 4]]];
```

In[77]:= `(* Attach an external propagator to the ball with a w2 vertex *)`

In[78]:= `L[a_, b_, c_, d_] = Expand[K[a, b] G[a, b, c, d]];`

In[79]:= ```
F2 = w2 Simplify[prod[G[1, 2, 1, 2], G[1, 2, 1, 3],
     G[1, 2, 3, 4], L[1, 2, 1, 2], L[1, 2, 1, 3], L[1, 2, 3, 4]]];
```

In[80]:= `(* Total ball diagram *)`

In[81]:= ```
Ftot = Simplify[Simplify[F1 + F2] /. invk1 /. inv /. n → 1];
(* This is the propagator correction up to a q integration *)
```

In[82]:= `Fser = Simplify[Series[Ftot, {m1, 0, -4}]][[1]][[1]]`

Out[82]= $\left\{ \dfrac{32\,(m2+m3)^2\,(w1-w2)^2}{(1+q^2)^2\,m1^6} + \dfrac{8\,(m2+m3)\,(w1-w2)\,\left(4\,m3\,w1 + m2\,\left((13+6\,q^2)\,w1 - (3+2\,q^2)\,w2\right)\right)}{m2\,(1+q^2)^2\,m1^5} + \right.$

$\dfrac{1}{m2^2\,(1+q^2)^2\,m1^4}\,4\,\Big(16\,m3^2\,w1\,(w1-w2)\, +$

$\quad 4\,m2\,m3\,\left((20+15\,q^2+6\,q^4)\,w1^2 - (22+21\,q^2+10\,q^4)\,w1\,w2 + 4\,(1+q^2)^2\,w2^2\right) +$

$\quad m2^2\,\left((73+69\,q^2+24\,q^4)\,w1^2 - 2\,(39+45\,q^2+20\,q^4)\,w1\,w2 + (17+33\,q^2+16\,q^4)\,w2^2\right)\Big) +$

$\dfrac{1}{O[m1]^3}$, $\dfrac{32\,(m2+m3)^2\,(w1-w2)^2}{(1+q^2)^2\,m1^6} + \dfrac{8\,(m2+m3)\,\left(4\,m3\,w1 + m2\,(5+2\,q^2)\,(3\,w1-w2)\right)\,(w1-w2)}{m2\,(1+q^2)^2\,m1^5} +$

$\dfrac{1}{m2^2\,(1+q^2)^2\,m1^4}\,4\,\Big(16\,m3^2\,w1\,(w1-w2)\, +$

$\quad 4\,m2\,m3\,\left((22+15\,q^2+6\,q^4)\,w1^2 - (24+21\,q^2+10\,q^4)\,w1\,w2 + 4\,(1+q^2)^2\,w2^2\right) + m2^2$

$\quad \left(3\,(29+25\,q^2+8\,q^4)\,w1^2 - 2\,(47+49\,q^2+20\,q^4)\,w1\,w2 + (19+35\,q^2+16\,q^4)\,w2^2\right)\Big) + \dfrac{1}{O[m1]^3}$,

$\dfrac{32\,(m2+m3)^2\,(w1-w2)^2}{(1+q^2)^2\,m1^6} + \dfrac{16\,(m2+m3)\,(w1-w2)\,\left(2\,m3\,w1 + m2\,\left((8+3\,q^2)\,w1 - (3+q^2)\,w2\right)\right)}{m2\,(1+q^2)^2\,m1^5} +$

$\dfrac{1}{m2^2\,(1+q^2)^2\,m1^4}\,8\,\Big(8\,m3^2\,w1\,(w1-w2)\, +$

$\quad 2\,m2\,m3\,\left((23+15\,q^2+6\,q^4)\,w1^2 - (25+21\,q^2+10\,q^4)\,w1\,w2 + 4\,(1+q^2)^2\,w2^2\right) +$

$\quad \left. m2^2\,\left((47+39\,q^2+12\,q^4)\,w1^2 - (51+51\,q^2+20\,q^4)\,w1\,w2 + 2\,(5+9\,q^2+4\,q^4)\,w2^2\right)\Big) + \dfrac{1}{O[m1]^3} \right\}$

In[83]:= `Simplify[Fser /. w2 → w1]`

Out[83]= $\left\{ \dfrac{16\,(3\,m2+2\,m3)\,w1^2}{m2\,(1+q^2)\,m1^4} + \dfrac{1}{O[m1]^3},\ \dfrac{16\,(3\,m2+2\,m3)\,w1^2}{m2\,(1+q^2)\,m1^4} + \dfrac{1}{O[m1]^3},\ \dfrac{16\,(3\,m2+2\,m3)\,w1^2}{m2\,(1+q^2)\,m1^4} + \dfrac{1}{O[m1]^3} \right\}$

In[84]:= `(* We see that also for this term the result`
`is O(1/m1^4) and therefore irrelevant in dimension 6 *)`


\end{document}